\documentclass{article}

\usepackage{arxiv}
\usepackage{amsmath, amssymb, amsfonts, mathtools}
\usepackage{multirow}
\usepackage[utf8]{inputenc} % allow utf-8 input
\usepackage[T1]{fontenc}    % use 8-bit T1 fonts
\usepackage{hyperref}       % hyperlinks
\usepackage{url}            % simple URL typesetting
\usepackage{booktabs}       % professional-quality tables
\usepackage{amsfonts}       % blackboard math symbols
\usepackage{nicefrac}       % compact symbols for 1/2, etc.
\usepackage{microtype}      % microtypography
\usepackage{lipsum}		% Can be removed after putting your text content
\usepackage{graphicx}
\usepackage{natbib}
\usepackage{doi}
\usepackage{setspace}

\title{Bayesian Joint Modeling For Longitudinal Magnitude Data With Informative Dropout: An Application To Critical Care Data}

\date{} 					% Or removing it

\author{Wen Teng$^{1,*}$, Niall D.~Ferguson$^{2}$, Ewan C.~Goligher$^{2}$ and Anna Heath$^{1,3,4}$ \vspace{0.25em}\\
$^{1}$ Child Health Evaluative Sciences, The Hospital for Sick Children, Toronto, ON, Canada \vspace{0.25em}\\
$^{2}$ Department of Medicine, Division of Respirology, University Health Network, Toronto, ON, Canada \vspace{0.25em}\\
$^{3}$ Division of Biostatistics, Dalla Lana School of Public Health, University of Toronto, Toronto, ON, Canada \vspace{0.25em}\\
$^{4}$ Department of Statistical Science, University College London, London, U.K.\vspace{0.25em}\\
$*$email:\ \texttt{wen.teng@sickkids.ca}}

\renewcommand{\headeright}{}
\renewcommand{\undertitle}{}
\renewcommand{\shorttitle}{Bayesian Joint Modeling for Longitudinal Magnitude Data}

\hypersetup{
pdftitle={A template for the arxiv style},
pdfsubject={q-bio.NC, q-bio.QM},
pdfauthor={David S.~Hippocampus, Elias D.~Striatum},
pdfkeywords={First keyword, Second keyword, More},
}

\begin{document}
\maketitle

\begin{abstract}
In various biomedical studies, analysis often focuses on data magnitudes, particularly when algebraic signs are irrelevant or lost. 
For repeated measures studies involving magnitude outcomes, incorporating random effects is essential as they account for individual heterogeneity, thereby enhancing parameter estimation precision. 
However, established regression methods specifically designed for magnitude outcomes that incorporate random effects are currently lacking.
This article bridges this gap by introducing Bayesian regression modeling approaches for analyzing magnitude data, with a key focus on incorporating random effects. 
The proposed method is further extended to address multiple causes of informative dropout, a common challenge in repeated measures studies. 
To tackle this missing data challenge, a joint modeling strategy is developed, building upon the introduced regression techniques.
Two numerical simulation studies assess the validity of our method. 
The chosen simulation scenarios are designed to resemble the conditions of our motivating study. 
Results demonstrate that the proposed method for magnitude data performs well in terms of estimation accuracy, and the joint models effectively mitigate bias due to missing data.
Finally, we apply these models to analyze magnitude data from the motivating study, investigating whether sex impacts the magnitude change in diaphragm thickness over time for ICU patients.
\end{abstract}

% keywords can be removed

\section{Introduction}
Across a range of biomedical and industrial studies, researchers often model the magnitude of a particular quantity \citep{Liu2020}. 
For instance, clinical studies might measure health or organ function as an absolute deviation from a baseline or reference standard \citep{goligher2015evolution}. 
Similarly, in communication theory, the signal-to-noise ratio, a key magnitude measure, is crucial for assessing electronic devices \citep{freeman1958principles}. 
In other contexts, only the outcome's magnitude may be measurable. Consequently, the outcome of interest is characterized by a distribution of magnitudes, rather than its own distribution. 
Geometrically, this can be visualized as the negative side of the distribution being ``folded'' onto the positive side.

Where the underlying continuous outcome is assumed to follow a normal distribution, its magnitude can be modeled as a folded normal distribution \citep{Leone1961}. 
The concept of the folded normal distribution was initially introduced in the 1960s by \citet{Leone1961}, who also formulated a moment estimator method for parameter estimation. 
\citet{Elandt1961} later significantly contributed by proposing an alternative estimation approach. 
Building on these foundational works, subsequent studies have further enriched our understanding of the folded normal distribution \citep{Johnson1962,Sundberg1974O,Tsagris2014}.

Quantifying the relationship between predictors and study outcomes often requires regression approaches. 
Nevertheless, research on folded normal regression is scarce. 
To our knowledge, only one article examines regression methods with a folded normal response, and it is limited to fixed effects \citep{Liu2020}. 
Our research focuses on repeated measures studies, which necessitate mixed-effects models with random effects to account for both within-subject and between-subject variability. 
However, no established regression methods currently exist for a folded normal response that incorporate both fixed and random effects. 
Therefore, we developed a folded normal regression method to address this gap.

A significant challenge in introducing random effects into folded normal regression is that it leads to a non-identifiable model \citep{lehmann2006}. 
Generally, a model is ``identifiable'' when distinct parameter values result in different distributions of observations; this criterion is not met by folded normal models \citep{Leone1961} unless parameter constraints are applied. 
However, incorporating random effects demands more complex constraints.
Therefore, to address this non-identifiability challenge in the context of analyzing folded normal data from repeated measures studies, this article introduces a Bayesian modeling approach, leveraging its flexibility to incorporate random effects and ensure identifiability.

Another key issue in repeated measures studies is dropout that can be related to the underlying value of the outcome, known as informative or nonignorable dropout \citep{little1995,yuan2009mixed}. 
This leads to data typically characterized as ``missing not at random'', which means that ignoring missingness due to informative dropout can introduce bias into study results and conclusions \citep{geskus2014individuals}. 
Therefore, to ensure accurate statistical inference, our models must be extended to address the potential bias caused by missing data.

The issue of missing data in longitudinal studies with informative dropout has been extensively explored using likelihood-based methods, as comprehensively reviewed by \citet{little1995}, \citet{molenberghs2007}, and \citet{daniels2008}. 
These models vary in how they factorize the joint distribution of outcome and missing data mechanisms, including selection models \citep{heckman1979,wu1988,diggle1994}, pattern-mixture models \citep{wu1989,little1993,little1994}, shared parameter models \citep{wu1988estimation,de1994modelling,pulkstenis1998model}, and mixed-effects hybrid models \citep{yuan2009mixed,ahn2013bayesian}.

In our study, we handle missing data caused by informative dropout using likelihood-based methods to model the underlying mechanism. 
In our applied context, informative dropouts arise from multiple causes. 
Thus, our model for the dropout process accounts for these complexities by utilizing discrete time survival models \citep{suresh2022survival} to construct competing risk models. 
This approach treats dropouts with different causes as distinct types of events, acknowledging that one type of dropout precludes others.

Overall, we adopt a joint modeling strategy that combines folded normal responses with multiple causes of informative dropouts. 
This approach simultaneously models the magnitude outcome process and the dropout process, linking them through shared subject-specific random effects. 
This facilitates analyzing magnitude data from repeated measures studies, particularly when responses are missing due to diverse causes of informative dropouts.

The article is organized as follows: Section 2 introduces the motivation of this study. 
In Section 3, we present and discuss the key elements of our proposed statistical models. 
Section 4 details our simulation studies, evaluating model performance against standard regression. 
Section 5 applies our methods to data from the motivating study. 
Finally, Section 6 provides conclusions.

\section{Motivation}
\begin{figure}[htbp]
  \centering
  \includegraphics[width= 0.9\textwidth]{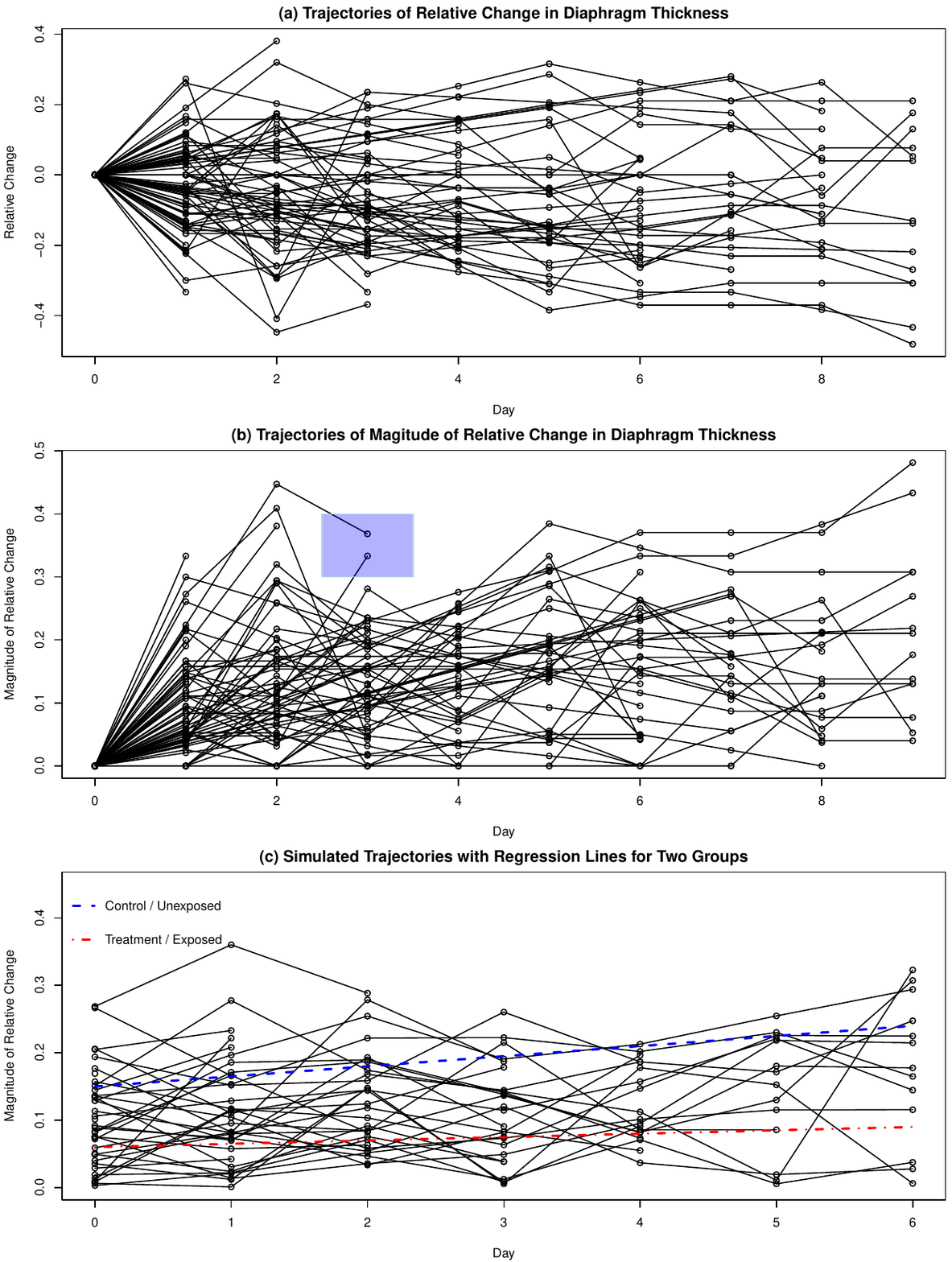} % Adjust width as needed
  \caption{(a) Individual trajectories of relative change in diaphragm thickness over 9 days (Day 1-9, with Day 0 as baseline) for 100 randomly selected subjects from the motivating study. (b) Magnitude of relative change in diaphragm thickness for the same subjects, with a shaded rectangle illustrating informative dropout. (c) Simulated individual trajectories of magnitude of relative change for 40 subjects over 7 days (Day 0-6), with their true underlying regression lines: Treatment/Exposed group (dot-dash line) and Control/Unexposed group (dashed line).}
  \label{fig:plot}
\end{figure}

The methodology presented in this article was developed to address challenges encountered with data from the ``Evolution of Diaphragm Thickness during Mechanical Ventilatio'' study \citep{goligher2015evolution}. 
This clinical study investigated daily changes in diaphragm thickness in critically ill patients receiving mechanical ventilation. 
Clinically, interest lies in the absolute change in diaphragm thickness, as both increases and decreases from baseline are associated with adverse outcomes and are considered equivalent in magnitude. 
Therefore, the focus is on measuring the magnitude of change rather than its direction.
Figure \ref{fig:plot}, panels (a) and (b), show individual trajectories of relative change and its magnitude in diaphragm thickness for 100 randomly selected subjects from the motivating study, respectively. 
Panel (b) is obtained by folding the negative side of panel (a) onto its positive side.

The clinical team is particularly interested in understanding the association between the magnitude of change in diaphragm thickness and a specific exposure, and aims to quantify this association as the average difference in the magnitude of change over time between exposed and unexposed groups.
To address this, we developed a regression method designed to accurately model this average difference (the exposure effect) while appropriately handling the magnitude of change as the outcome. 
Our approach specifically uses the average distance between regression lines for exposed and unexposed groups to represent the average difference.
We employ a mixed-effects model to adjust for the repeated measurements taken from individual patients. 
Furthermore, the error term in our model is based on the folded normal distribution, as a standard normal error is not suitable  when absolute changes in diaphragm thickness are the outcome.
Panel (c) of Figure \ref{fig:plot} shows a simulated example of 40 subjects in two groups, whose magnitude of change in diaphragm thickness is generated by regression lines (dot-dash and dashed lines) in each group.
The errors in this example are not normally distributed, and instead, are set to follow a folded normal distribution. 
Our objective of inference is the accurate estimation of the average distance between these two regression lines.

The motivating study also exhibits informative dropout due to multiple causes.
Approximately $80\%$ of patients lack diaphragm thickness measurements during their nine days of ventilation. 
This dropout occurs either when a patient recovers and no longer requires mechanical ventilation, or when the patient dies.
Crucially, diaphragm thickness is associated with the chance of both recovery and death; this means these dropouts are informative and could significantly bias estimates of the exposure effect.
The shaded rectangle of panel (c) of Figure \ref{fig:plot} highlights two cases of dropout.
To handle this challenge, we propose a joint modeling approach that incorporates the missing data mechanism directly into the previously described folded normal mixed-effects model.
This joint model is designed to mitigate the bias caused by informative dropout.

Beyond our motivating study, the magnitude of change in diaphragm thickness is a crucial endpoint in critical care trials for mechanical ventilation. 
With interventional studies to improve critical care outcomes becoming increasingly common, as exemplified by the ongoing LANDMARK II trial explicitly including this as its key efficacy outcome, accurate modeling methods for this endpoint hold broad clinical and research importance.
%Interventional studies to improve critical care outcomes are increasingly common, especially with the COVID-19 pandemic and the resulting influx of intensive care patients. 
%For instance, the ongoing LANDMARK II trial explicitly includes this endpoint as its key efficacy outcome. 
%Thus, developing modeling methods to provide accurate estimates for this endpoint holds broad clinical and research importance beyond our specific motivating study.

\section{Methods}
This section introduces the regression models used to provide reliable statistical inference for continuous magnitude data. We initially introduce the the folded normal distribution and review the folded normal fixed-effects regression methods. 
Subsequently, we propose a novel mixed-effects folded normal regression model. Finally, we account for missing data due to informative dropout by introducing joint models that consider outcome and dropout processes simultaneously, using shared random effects.
These models are suitable for observational and interventional studies estimating an exposure or treatment effect. 
For simplicity, we use ``exposure effect'' exclusively as these study types are equivalent from a modeling perspective.

\subsection{Folded Normal Distribution} 
Continuous measurements are often assumed to follow a normal distribution, denoted as, $Y \sim \mathcal{N}(\mu, \sigma^2)$ , where $\mu$ and $\sigma^2$ are the mean and variance of the normal distribution $\mathcal{N}(\mu, \sigma^2)$.
The magnitude of $Y$, $Z = |Y|$, then follows a folded normal distribution with probability density function
\begin{align*}
    f_{\mathcal{FN}}(z\mid\mu, \sigma^2) &= \frac{1}{\sqrt{2\pi}\sigma}\left[ \exp\left\{-\frac{\left(z-\mu\right)^2}{2\sigma^2}\right\} + \exp\left\{-\frac{\left(z+\mu\right)^2}{2\sigma^2}\right\} \right]\\
    &= \mathcal{N}(z\mid\mu, \sigma^2) + \mathcal{N}(-z\mid\mu, \sigma^2), \quad z \ge 0,
\end{align*}
denoted as $Z \sim \mathcal{FN}(\mu, \sigma^2)$, where $\mathcal{N}(z\mid\mu, \sigma^2)$ is the density of a normal distribution $\mathcal{N}(\mu, \sigma^2)$.

When considering the folded normal distribution, it is crucial to acknowledge that
\begin{align*}
f_{\mathcal{FN}}(z\mid\mu, \sigma^2) = f_{\mathcal{FN}}(z\mid-\mu, \sigma^2).
\end{align*}
This means that two distinct parameter values, $\pm\mu$ yield the same density, causing a non-identifiability problem.
This complicates likelihood-based estimation of $\mu$ since observed data cannot uniquely determine its true value. 
This issue is easily resolved by restricting $\mu \ge 0$, for instance, by using a prior distribution with non-negative support in a Bayesian framework.

Another key feature of the folded normal distribution is that, when $\mu > 0$ and $\sigma$ is small relative to $\mu$, the difference between $\mathcal{N}(\mu, \sigma^2)$ and $\mathcal{FN}(\mu, \sigma^2)$ becomes negligible in $[0, +\infty)$. 
In this scenario, $\mu$ estimates are likely similar, regardless of whether a folded normal or normal model is assumed.

\subsection{Review of Folded Normal Fixed-Effects Model}
\citet{Liu2020} extended the folded normal distribution by constructing a fixed-effects regression model.
This approach models independent random variables $Z_{i}$, for $i = 1, 2, \ldots, N$, following $\mathcal{FN}(\mu_i, \sigma^2)$, with $\mu_i$ defined as a linear predictor:
\begin{align*}
Z_{i}\mid\textbf{X}_i, \textbf{B}  \sim \mathcal{FN}(\mu_{i}, \sigma^2), \quad
\mu_{i} = \textbf{X}_i^\top \textbf{B},
\end{align*}
where $\textbf{X}_i$ and $\textbf{B}$ represent a covariate vector for individual $i$ and a corresponding coefficient vector.
Our regression model adopts this concept while adding random effects and extra constraints.

\citet{Liu2020} estimated model parameters using both frequentist (EM algorithm \citep{Dempster1977} with bootstrap for confidence intervals) and Bayesian (data augmentation \citep{tanner1987calculation} for posterior samples) methods. 
Our proposed inference, detailed below, uses standard Bayesian modeling, differing from their approach.

\subsection{Folded Normal Mixed-Effects Model}
Leveraging the knowledge of folded normal distribution and the fixed-effects model introduced by \citet{Liu2020}, we develop a folded normal mixed-effects model for repeated measures studies. 

\subsubsection{Notation} 
We start by outlining some notation for our methodology.
In a longitudinal study, we have $K$ repeated measurements taken at times $t_1, t_2, \ldots, t_K$. 
To simplify the modeling, we set $t_k = k - 1$ for $k = 1, 2, \ldots, K$.
For each subject $i$, $x_i$ is a binary exposure indicator (0 for unexposed group, 1 for exposed group).
The measurement value for subject $i$ at time $t$ is $Y_{it}$ and our response of interest is its magnitude, $Z_{it} = |Y_{it}|$.
Thus, the complete set of observations for subject $i$ is $\textbf{Z}_i = \left(Z_{it}\right)_{t = 0}^{K-1}$.

\subsubsection{Modeling}
This study uses linear curves to model the relationship with time, consistent with our motivating study, where measurement magnitudes exhibited generally linear trajectories over time.
Our modeling goal is to quantify the average difference in the true magnitude outcomes between exposed and unexposed groups (exposure effect). 
This is defined as the average distance $AD$ between the two exposure groups' regression lines across times $t_1, t_2, \ldots, t_K$.
Constructing mixed-effects models with folded normal magnitude responses is crucial for a reliable $AD$ estimate.

For each subject $i$, we introduce random effects denoted as $\textbf{b}_i$, to account for the correlation among repeated measurements and to accommodate individual variability in response over time.
Our Bayesian inference relies on the joint distribution of $\textbf{Z}_i$ and $\textbf{b}_i$ given the exposure indicator $x_i$ and model parameters $\boldsymbol\phi$, denoted as $f(\textbf{Z}_i, \textbf{b}_i\mid x_i, \boldsymbol\phi)$.
For subject $i$, we use the following factorization:
\begin{align*}
    f(\textbf{Z}_i, \textbf{b}_i\mid x_i, \boldsymbol\phi) = f(\textbf{b}_i\mid x_i, \boldsymbol\phi_1)
    f(\textbf{Z}_i\mid\textbf{b}_i, x_i, \boldsymbol\phi_2),
\end{align*}
where $\boldsymbol\phi = \left(\boldsymbol\phi_1, \boldsymbol\phi_2\right)^\top$, with $\boldsymbol\phi_1$ and $\boldsymbol\phi_2$ representing the parameters for the random effects density and the conditional density of outcomes given random effects, respectively.
%The first component of the right-hand side of (\ref{f1}) describes the marginal distribution of random effects;
%the second component models the outcome process, conditioned on the random effects. 

\subsubsection{Likelihood}
More specifically, the outcome process for subject $i$ at times $t= 0, 1, \dots, K-1$ represented by  $f(\textbf{Z}_i\mid\textbf{b}_i, x_i, \boldsymbol\phi_2)$, is specified as follows:
\begin{align}\label{f2}
\begin{split}
&Z_{it}\mid\textbf{b}_i, x_i, \boldsymbol\phi_2 \sim \mathcal{FN}(\mu_{it}, \sigma^2), \\
&\mu_{it} = 
\begin{cases}
c_0 + \alpha_{0i} + \left(c_1 + \alpha_{1i}\right) t, &\  x_i = 0 \\
d_0 + \beta_{0i} + \left(d_1 + \beta_{1i}\right) t, &\ x_i = 1
\end{cases}
\end{split}
\end{align}
where $c_0$, $c_1$, $\alpha_{0i}$ and $\alpha_{1i}$ are the intercept, slope, random intercept and random slope for the unexposed group, respectively, 
and $d_0$, $d_1$, $\beta_{0i}$ and $\beta_{1i}$ are those for the exposed group.
Moreover, $\textbf{b}_i = \left(\alpha_{0i}, \alpha_{1i}, \beta_{0i}, \beta_{1i}\right)^\top$ is the random effect vector, and $\boldsymbol\phi_2 = \left(c_0, c_1, d_0, d_1, \sigma^2\right)^\top$ is the model parameter vector.
In model (\ref{f2}), $\mu_{it}$ represents the true underlying trajectory of magnitude outcomes (with folded normal error excluded) over time for subject $i$.
In the unexposed group, the mean subject trajectory, or regression line, is $c_0 + c_1 t$. 
For the exposed group, it is $d_0 + d_1 t$.
%To prevent non-identifiability, the majority of true trajectories $\mu_{it}$ should be non-negative. 
Since the true trajectories $\mu_{it}$ are non-negative, the distinct regression lines between the exposure groups indicate that the random intercepts ($\alpha_{0i}$ and $\beta_{0i}$) and random slopes ($\alpha_{1i}$ and $\beta_{1i}$) follow different distributions between groups. 
This reflects variations around each group's mean trajectory.
Note that, our primary target of inference, the average distance $AD$, can be calculated from model (\ref{f2}) as follows:
\begin{align*}
AD = \frac{1}{K}\sum_{t=0}^{K-1}\left[c_0 + c_1  t - \left(d_0 + d_1 t\right)\right] = c_0 - d_0 + \left( c_1 - d_1 \right) \frac{K-1}{2}.
\end{align*}

The model for the process described by random effects $\textbf{b}_i$, 
represented by $f(\textbf{b}_i\mid x_i, \boldsymbol\phi_1)$, is then:
\begin{align}\label{f3}
\left(\alpha_{0i}, \alpha_{1i}\right)^\top \mid \Sigma_\alpha \sim \mathcal{N}_2(\textbf{0}, \Sigma_\alpha),\quad \left(\beta_{0i}, \beta_{1i}\right)^\top \mid \Sigma_\beta \sim \mathcal{N}_2(\textbf{0}, \Sigma_\beta)
\end{align}
with
\begin{align}\label{f3-2}
\Sigma_\alpha = 
\begin{pmatrix} \tau_{\alpha0}^2 & \rho_\alpha\tau_{\alpha0}\tau_{\alpha1} \\ \rho_\alpha\tau_{\alpha0}\tau_{\alpha1} & \tau_{\alpha1}^2 \end{pmatrix},
\quad
\Sigma_\beta = 
\begin{pmatrix} \tau_{\beta0}^2 & \rho_\beta\tau_{\beta0}\tau_{\beta1} \\ \rho_\beta\tau_{\beta0}\tau_{\beta1} & \tau_{\beta1}^2 \end{pmatrix},
\end{align}
where 
$\boldsymbol\phi_1 = (\tau_{\alpha0}, \tau_{\alpha1}, \rho_\alpha, \tau_{\beta0}, \tau_{\beta1}, \rho_\beta)^\top$ is the model parameter vector. 
Here, $\tau_{\alpha0}$, $\tau_{\alpha1}$, $\tau_{\beta0}$ and $\tau_{\beta1}$ are the standard deviations of their corresponding random effects.
$\rho_\alpha$ and $\rho_\beta$ represent the correlations between the random intercept and random slope in the unexposed and exposed groups, respectively. 
In this model, the random effects within each group follow independent bivariate normal distributions, accounting for the relationship between random intercepts and random slopes.

\subsubsection{Priors and Constraints}
To address non-identifiability and enforce non-negativity of the regression lines in model (\ref{f2}), we introduce constraints through the following priors:
\begin{align}\label{f4}
\begin{split}
&c_0, c_1, d_0, d_1 \sim \mathcal{TN}(0, \tilde{\mu}, \tilde{\sigma}^2 ), \\
&\tau_{\alpha0}\mid c_0, \omega \sim \mathcal{U}(0, c_0/\omega), \quad \tau_{\alpha1}\mid c_1, \omega \sim \mathcal{U}(0, c_1/\omega), \\
&\tau_{\beta0}\mid d_0, \omega \sim \mathcal{U}(0, d_0/\omega), \quad \tau_{\beta1}\mid d_1, \omega \sim \mathcal{U}(0, d_1/\omega), \\
&\omega \sim \mathcal{U}(b_L, b_U), \\
&\rho_\alpha, \rho_\beta \sim \mathcal{U}(-1, 1),
\end{split}
\end{align}
where $\mathcal{TN}(0, \tilde{\mu}, \tilde{\sigma}^2 )$ is the truncated normal distribution, obtained by truncating the normal distribution $\mathcal{N}(\tilde{\mu}, \tilde{\sigma}^2)$ from below, with the truncation threshold set to 0, and $\mathcal{U}(a, b)$ is the uniform distribution on the interval $(a, b)$ for real numbers $a < b$.

The first row of (\ref{f4}) enforces non-negative intercepts ($c_0$ and $d_0$) and slopes ($c_1$ and $d_1$) for each group's regression lines, a crucial condition for identifiability consistent with our motivating study's predominantly increasing absolute change from baseline. 
Note that $\tilde{\mu}$ and $\tilde{\sigma}^2$ need not be identical for each fixed effect if prior information is available.
The next three rows of (\ref{f4}) address non-identifiability by constraining the standard deviations of the random effects (the $\tau$ parameters) through uniform priors.
These priors are bounded by fixed effects and $\omega$, where $\omega$ is learned from data via its own uniform prior over $(b_L, b_U)$.
This structure ensures estimated individual true trajectories are highly likely to be non-negative, which, in turn, enables fixed effect identification and valid estimates.
Finally, the last rows specify that correlations are constrained to lie between -1 and 1.

We establish minimally informative priors by setting $\tilde{\mu}$ as a constant near zero and $\tilde{\sigma}^2$ as a large, non-negative constant.
From our experience, the selection of $b_L$ and $b_U$ is quite flexible; we recommend $b_L \le 2 \le b_U$ based on the ``two-sigma rule'' \citep{upton2014dictionary} and, specifically, set $b_L = 1.2$ and $b_U = 2.4$.
In the outcome model (\ref{f2}), we use a minimally informative inverse gamma distribution for $\sigma^2$.

\subsubsection{Implementation}
The implementation of our proposed model requires non-centered parameterization with Cholesky decomposition \citep{golub2013matrix} of the covariance matrices (\ref{f3-2}).
This approach is essential to ensure the convergence of the Markov chains for model parameters' samples, as demonstrated by our extensive practice.
We also observed that Stan performed significantly worse than JAGS, suggesting our model's posterior geometry might be challenging for gradient-based methods (HMC/NUTS) compared to JAGS's methods (Gibbs/Metropolis).

Overall, models (\ref{f2}), (\ref{f3}) and (\ref{f3-2}), along with constraints (\ref{f4}), form a complete Bayesian model for longitudinal magnitude outcomes without missing data. 
We will now extend this to our joint modeling framework to account for informative dropout.

\subsection{Joint Model for Informative Dropout}

\subsubsection{Notation}
Our motivating study includes missingness due to multiple causes of informative dropout. 
Given this, we assume that the underlying data includes $H$ mutually exclusive reasons for informative dropout. 
We further assume every subject has at least one observed response, as individuals without observations provide no information and are thus excluded from the analysis.
Consequently, $D$ is defined as the last observed measurement time before dropout or the end of observation period, and $\delta$ indicates the reason for dropout, where $\delta = 0, \dots, H$ ($0$ representing no dropout). 
If subject $i$ experiences dropout between measurement times $t$ and $t+1$ due to the $h$-th cause, this is denoted as $D_i = t$ and $\delta_i = h$, with $t$ taking values from from $t_1 = 0$ to $t_{K-1} = K-2$.
If subject $i$ completes follow-up, it is denoted as $D_i = t_K$ and $\delta_i = 0$. 
The observed measurement values and magnitude responses for subject $i$ are then $\textbf{Y}_i^o = \left( Y_{it} \right)_{t = 0}^{D_i}$ and $\textbf{Z}_i^o = \left( Z_{it} \right)_{t = 0}^{D_i}$, respectively.

\subsubsection{Modeling}
Our aim is to appropriately model magnitude responses despite missing data caused by informative dropout.
We achieve this using a joint modeling approach that simultaneously models both the observed outcome process and the missing data mechanism.
Within this framework, the likelihoods of the outcome and dropout processes are linked via shared random effects.

Specifically, the joint model is constructed based on the joint distribution of $\textbf{Z}_i^o$, $D_i$, $\delta_i$, and $\textbf{b}_i$ for subject $i$ given the exposure indicator $x_i$ and model parameters $\boldsymbol\theta$, denoted as $f(\textbf{Z}_i^o, D_i, \delta_i, \textbf{b}_i\mid x_i, \boldsymbol\theta)$. 
We factorize this joint distribution into three components:
\begin{align}\label{f5}
    f(\textbf{Z}_i^o, D_i, \delta_i, \textbf{b}_i\mid x_i, \boldsymbol\theta) = f(\textbf{b}_i\mid x_i, \boldsymbol\theta_1)f(\textbf{Z}_i^o\mid \textbf{b}_i, x_i, \boldsymbol\theta_2)f(D_i, \delta_i\mid \textbf{b}_i, x_i, \boldsymbol\theta_3),
\end{align}
where $\boldsymbol\theta = (\boldsymbol\theta_1, \boldsymbol\theta_2, \boldsymbol\theta_3)$.
Here, $\boldsymbol\theta_1$, $\boldsymbol\theta_2$ and $\boldsymbol\theta_3$ represent the parameters for the density of random effects, the conditional density of observed outcomes given random effects, and the conditional density of dropout patterns given random effects, respectively.

The product $f(\textbf{b}_i\mid x_i, \boldsymbol\theta_1)f(\textbf{Z}_i^o\mid \textbf{b}_i, x_i, \boldsymbol\theta_2)$ on the right-hand side of equation (\ref{f5}) is addressed using our proposed folded normal regression method from the previous section.
The key difference in the current application is that this method now models the observed outcomes, $\textbf{Z}_i^o$, rather than the complete data, $\textbf{Z}_i$.
The remaining term, $f(D_i, \delta_i\mid \textbf{b}_i, x_i, \boldsymbol\theta_3)$, pertains to the dropout mechanism.
This term is incorporated based on the reasonable assumption that the likelihood of subject $i$ dropping out at time $t$ depends on its true magnitude outcomes at that same time (i.e., $\mu_{it}$ in model (\ref{f2})),  which, in turn, relies on $\textbf{b}_i$ and $x_i$.
We will now proceed to formulate a model specifically for this dropout term.

\subsubsection{Dropout Submodel}
For simplicity, we present our method in the case of two causes of informative dropout, i.e., $H = 2$. 
This aligns with our motivating study, where recovery and death are the distinct causes of dropout.
For subject $i$ at measurement time $t$ where $t_1 \le t < t_K$, we define the hazards of type 1 and type 2 dropout, respectively, as
\begin{align*}
    \lambda_{it} = \text{Pr}(D_i = t, \delta_i = 1 \mid D_i \ge t, \textbf{b}_i, x_i), \\
    \kappa_{it} = \text{Pr}(D_i = t, \delta_i = 2 \mid D_i \ge t, \textbf{b}_i, x_i).
\end{align*}

Using these hazards, the model for the $h$-th dropout is constructed in two parts.
The first part employs a generalized logit link for regression, capturing the relationship between the hazards and the linear predictors of dropout model:
\begin{align}\label{f6}
\begin{split}
&\log\left(\frac{\lambda_{it}}{1-\lambda_{it}-\kappa_{it}}\right) = 
\begin{cases}
q_0(t) + p_0 \mu_{it}, &\  x_i = 0 \\
q_1(t) + p_1 \mu_{it}, &\ x_i = 1
\end{cases}\\
&\log\left(\frac{\kappa_{it}}{1-\lambda_{it}-\kappa_{it}}\right) = 
\begin{cases}
v_0(t) + u_{0} \mu_{it}, &\  x_i = 0 \\
v_1(t) + u_{1} \mu_{it}, &\ x_i = 1
\end{cases}\end{split}
\end{align}
where parameters $p_0$, $p_1$, $u_0$ and $u_1$ quantify the association between the true underlying longitudinal magnitude outcomes ($\mu_{it}$, as defined in model (\ref{f2})) and the dropout hazards. 
Meanwhile, $q_0(t)$, $q_1(t)$, $v_0(t)$ and $v_1(t)$ are functions that govern the relationship between measurement time $t$ and dropout hazards, typically specified as:
\begin{align}\label{f7}
q_m(t) = \sum_{n = 0}^{K-2} q_{mn} I_{n}(t), \quad  v_m(t) = \sum_{n = 0}^{K-2} v_{mn} I_{n}(t), \quad m = 0, 1,
\end{align}
where $I_n(t)$ denotes the indicator function, equal to 1 when $t=n$ and 0 otherwise.
Equation (\ref{f7}) treats the measurement time as a categorical variable and are henceforth referred to as categorical time functions.
However, when the number of repeated measures $K$ is large, the functions in (\ref{f7}) would include an excessive number of parameters, leading to a significant computational burden.
One approach to reduce this burden is to combine measurement times and use an indicator variable for each combined level.
Alternatively, we can utilize the following linear functions:
\begin{align}\label{f8}
q_m(t) = q_{m0} + q_{m1}t, \quad  v_m(t) = v_{m0} + v_{m1}t, \quad m = 0, 1,
\end{align}
and these are termed linear time functions.
The model utilizing linear time functions requires significantly fewer parameters that the model using categorical time functions. 
In our simulation study, we will evaluate and compare their relative performance.

The second part of the model for type $h$ dropout is the likelihood of time to dropout:
\begin{align}\label{f9}
\begin{split}
&\text{Pr}(D_i = t, \delta_i = h \mid \textbf{b}_i, x_i, \boldsymbol\theta_3) \\
& = \lambda_{it}^{I_{1}(h)}\kappa_{it}^{I_{2}(h)}\prod\limits_{j=0}^{t-1}(1-\lambda_{ij}-\kappa_{ij}), \quad  0 \le t < K-1,\ h = 1, 2,\\
&\text{Pr}(D_i = K-1, \delta_i = 0\mid \textbf{b}_i, x_i, \boldsymbol\theta_3) = \prod\limits_{j=0}^{K-2}(1-\lambda_{ij}-\kappa_{ij}),
\end{split}
\end{align}
where $\boldsymbol\theta_3$ represents the model parameters, and $I_1(h)$ and $I_2(h)$ are indicator functions as previously defined. 
These likelihood functions collectively define a competing risk model for subject $i$.

The models defined in equations (\ref{f6}) and (\ref{f9}) provide a flexible regression structure capable of accommodating various dropout patterns. 
For their priors, we suggest that commonly used uninformative or minimally informative priors are sufficient.
In conclusion, by combining models (\ref{f2}), (\ref{f3})and (\ref{f3-2}) for observed outcomes with constraints (\ref{f4}), and models (\ref{f6}) and (\ref{f9}) for dropout patterns, we can effectively analyze magnitude data while accounting for multiple causes of informative dropout.

\section{Simulation Study}
This section details the numerical simulations used to evaluate our proposed methodology across various scenarios.
We first assess our folded normal regression model for complete data, followed by examining our joint models for handling informative dropout.
All simulations were conducted using Markov Chain Monte Carlo (MCMC) with JAGS 4.3.0 in R 4.0.2 environment, presented according to the framework described in \citet{morris2019using}.

\subsection{Folded Normal Mixed-Effects Model}

\subsubsection{Aims} 
This simulation study aims to compare our proposed mixed-effects model for folded normal responses with a standard linear mixed-effects model (the reference model).

\subsubsection{Data-generating Mechanisms} 
The data generating mechanisms in this simulation were chosen to closely mimic the data observed in the motivating study, rather than representing either the proposed or reference models.
For this simulation, 100 subjects were randomly assigned with a 0.5 probability to either the unexposed or exposed groups. 
Outcomes were measured at seven time points, $t_1, t_2, \ldots, t_7$, where $t_k = k-1$ for $k = 1, 2, \ldots, 7$.

The sign indicator $\gamma$ is a binary variable set to -1 with 0.6 probability and 1 with 0.4 probability.
This variable defines the sign of each subject's true measurement trajectories, indicating whether the natural scale outcome increases or decreases over time.
For subject $i$ at time $t$, in the unexposed group, measurement $Y_{it}$ is generated from a normal distribution $N(\gamma_i  \mu_{it}, \sigma^2)$ with $\mu_{it} = c_0 + \alpha_{0i} + \left(c_1 + \alpha_{1i}\right) t$, where $c_0$ and $c_1$ are fixed effects, and $\alpha_{0i}$ and $\alpha_{1i}$ are random effects.
Similarly, in the exposed group, the measurement $Y_{it}$ is generated from $N(\gamma_i \mu_{it}, \sigma^2)$ with $\mu_{it} = d_0 + \beta_{0i} + \left(d_1 + \beta_{1i}\right) t$, where $d_0$ and $d_1$ are fixed effects, and $\beta_{0i}$ and $\beta_{1i}$ are random effects.

We fixed $d_1$, $c_0$, and $c_1$ at 0.005, 0.15, and 0.015, respectively. 
We then varied $d_0$ across values of 0.08, 0.07, 0.06 and 0.05, which corresponded to increasing average distance ($AD$) values of 0.1, 0.11, 0.12 and 0.13. 
To assess model performance across different error variances, we set the standard deviation of the measurement error $\sigma$ to either 0.08 or 0.06.
Individual random effects $\alpha_{0i}$, $\alpha_{1i}$, $\beta_{0i}$, and $\beta_{1i}$, were generated from normal distributions: $N(0, (c_0/2)^2)$, $N(0, (c_1/2)^2)$, $N(0, (d_0/2)^2)$, and $N(0, (d_1/2)^2)$, respectively. 
Finally, $Z_{it}$, defined as the magnitude of $Y_{it}$, served as the response variable for both models.

\subsubsection{Estimands} 
We aim to estimate the average distance between the regression lines of the exposed and unexposed groups (exposure effect).
Based on the simulation parameters, this can be computed as $AD = c_0 - d_0 + 3(c_1 - d_1)$. Consequently, accurate estimates of $c_0$, $c_1$, $d_0$ and $d_1$ are required to estimate $AD$.

\subsubsection{Methods}
Our simulations compares our proposed folded normal mixed-effects model (defined in model (\ref{f2}), (\ref{f3}) and (\ref{f3-2}), and subject to the constraints (\ref{f4})), to a reference model. 
The reference model shares the same linear predictor, but assumes observed magnitude outcomes follow a normal distribution, thus eliminating the need for random effects constraints.
We conducted a factorial evaluation of the values for $d_0$ and $\sigma^2$ with 1000 simulation runs for each setting. 
In each simulation run, a burn-in of 2000 samples for both the folded normal and reference models ensured Markov chain convergence.
Subsequently, each model ran four chains, with 2000 samples per chain.

\subsubsection{Performance Measures}
Estimation of the average distance ($AD$) was evaluated by characterizing its posterior distribution. 
we computed the posterior mean, median, standard deviation, mean squared error, and the $2.5\%$ and $97.5\%$ quantiles. 
We also calculated the bias of the mean (the posterior mean minus the true $AD$), the relative bias (the bias divided by the true $AD$) and determined the containment of the true $AD$ within the $95\%$ credible interval (formed by these two quantiles).
Each of these posterior quantities was then averaged across all simulation runs to obtain corresponding overall measures, including the coverage probability derived from containment.
We additionally computed the empirical standard error of the posterior mean, defined as the sample standard deviation of the posterior means. 
Our primary performance measures for assessing accuracy were the average bias and average relative bias.

\subsubsection{Results} 
Table \ref{tab:Tab3} presents the simulation results. 
Overall, the average posterior mean closely aligns with the average posterior median for both models, indicating non-skewed distributions for $AD$. 
The empirical standard errors and average standard deviations are very similar, suggesting these models accurately characterize uncertainty.
The folded normal model tends to overestimate the average distance, resulting in a positive average bias, while the linear model exhibits a negative average bias. 
Importantly, in terms of absolute bias and relative bias, the folded normal model significantly outperforms the reference model across all scenarios, and its largest absolute relative bias is even smaller than the reference model's smallest.
This implies it can effectively eliminate the bias caused by errors following a folded normal distribution. 
This advantage is particularly evident when either the true $AD$ or $\sigma$ is large, aligning with our previous observation that the greatest differences between the folded normal and normal distributions occur when the standard deviation of the underlying normal distribution for $Y_{it}$ is large relative to its mean.
While the folded normal model yields slightly larger average standard deviations than the reference model (particularly when $\sigma$ is larger), its substantially smaller bias leads to coverage probabilities that align much better with the nominal level (within $1\%$).
This alignment ensures the accuracy and integrity of estimation, providing genuinely valid credible intervals, and indicating more reliable inference from the folded normal model.
Furthermore, the folded normal model exhibits smaller average mean squared errors than the reference model, reflecting its better overall performance in terms of both estimate accuracy and precision. 
Therefore, we conclude the folded normal model is superior to the reference model.

\begin{table}[ht]
	\renewcommand*{\arraystretch}{2}
	\caption{Performance comparison of folded normal (F) and linear (L) mixed-effects models across four true average distance (TAD) and two $\sigma$ values.}
	\centering
	\setlength{\tabcolsep}{4pt} % Reduces space between columns
	\begin{tabular}{ccccccccccccc}
		\hline 
		Model & $\sigma$ & TAD & Bias & Rel. Bias & Mean & Med. & S.D. & S.E. & $2.5\%$ & $97.5\%$ & C.P. & MSE \\ 
		\hline 
		\multirow{8}{1em}{F} & 0.06 & 0.10 & 0.00146 & 0.0146 & 0.1015 & 0.1014 & 0.0129 & 0.0125 & 0.0765 & 0.127 & 0.953 & 0.000327\\
        & 0.06 & 0.11 & 0.00154 & 0.0140 & 0.1115 & 0.1114 & 0.0126 & 0.0123 & 0.0870 & 0.137 & 0.951 & 0.000314\\
        & 0.06 & 0.12 & 0.00186 & 0.0155 & 0.1219 & 0.1217 & 0.0125 & 0.0120 & 0.0977 & 0.147 & 0.957 & 0.000305\\
        & 0.06 & 0.13 & 0.00256 & 0.0197 & 0.1326 & 0.1324 & 0.0123 & 0.0120 & 0.1087 & 0.157 & 0.956 & 0.000303\\
		& 0.08 & 0.10 & 0.00153 & 0.0153 & 0.1015 & 0.1014 & 0.0136 & 0.0131 & 0.0751 & 0.129 & 0.958 & 0.000361\\
        & 0.08 & 0.11 & 0.00193 & 0.0175 & 0.1119 & 0.1118 & 0.0135 & 0.0130 & 0.0859 & 0.139 & 0.956 & 0.000356\\
        & 0.08 & 0.12 & 0.00277 & 0.0231 & 0.1228 & 0.1226 & 0.0135 & 0.0131 & 0.0968 & 0.150 & 0.958 & 0.000362\\
        & 0.08 & 0.13 & 0.00442 & 0.0340 & 0.1344 & 0.1342 & 0.0138 & 0.0133 & 0.1080 & 0.162 & 0.953 & 0.000389\\
		\hline  
		\multirow{8}{1em}{L} & 0.06 & 0.10 & -0.00436 & -0.0436 & 0.0956 & 0.0956 & 0.0127 & 0.0122 & 0.0707 & 0.121 & 0.939 & 0.000330\\
        & 0.06 & 0.11 & -0.00553 & -0.0503 & 0.1045 & 0.1045 & 0.0124 & 0.0119 & 0.0801 & 0.129 & 0.927 & 0.000326\\
        & 0.06 & 0.12 & -0.00711 & -0.0593 & 0.1129 & 0.1129 & 0.0122 & 0.0116 & 0.0890 & 0.137 & 0.912 & 0.000335\\
        & 0.06 & 0.13 & -0.00917 & -0.0705 & 0.1208 & 0.1208 & 0.0120 & 0.0114 & 0.0973 & 0.144 & 0.887 & 0.000359\\
		& 0.08 & 0.10 & -0.00923 & -0.0923 & 0.0908 & 0.0908 & 0.0126 & 0.0121 & 0.0659 & 0.116 & 0.882 & 0.000392\\
        & 0.08 & 0.11 & -0.01138 & -0.1035 & 0.0986 & 0.0986 & 0.0124 & 0.0118 & 0.0743 & 0.123 & 0.848 & 0.000424\\
        & 0.08 & 0.12 & -0.01406 & -0.1172 & 0.1059 & 0.1059 & 0.0122 & 0.0116 & 0.0819 & 0.130 & 0.782 & 0.000484\\
        & 0.08 & 0.13 & -0.01730 & -0.1331 & 0.1127 & 0.1127 & 0.0121 & 0.0114 & 0.0889 & 0.137 & 0.694 & 0.000577\\
		\hline
	\end{tabular} 
	\label{tab:Tab3}
\end{table}

\subsection{Joint Model for Informative Dropout}

\subsubsection{Aims} 
This simulation study evaluates our proposed joint models for handling missing data, comparing them to a complete case analysis using our standard folded normal model as the reference.

\subsubsection{Data-generating Mechanisms} 
We initially simulate complete datasets using the model from the previous simulation study. 
We set the standard deviation of the measurement error $\sigma$ to 0.06. 
Consequently, we consider four simulation settings, corresponding to average distances $AD$ of 0.1, 0.11, 0.12 and 0.13. 

To mimic our motivating study, we consider two distinct causes of dropout: recovery and death.
These are represented by two survival models, one for time to recovery and another for time to death, which are then combined into a competing risk model. 
To define the informative dropout, we associate risk of dropout for individual $i$ with $R_i = \sum_{t=t_1}^{t_7} \mu_{it}/7$, the mean of its true trajectory of outcomes across all time points.

We denote time to recovery for subject $i$ as $T_i^r$ and time to death as $T_i^d$. 
The submodels for these times are: 
\begin{align*}
    T_i^r - 0.75 \sim \text{Gamma}(1 + 10  R_i,\, 50 R_i),
\end{align*}
\begin{align*}
    T_i^d \sim \text{Gamma}(1 + 0.5/R_i,\, 0.3/R_i).
\end{align*}
Here, $\text{Gamma}(k, \theta)$ is a Gamma distribution with shape parameter $k$, scale parameter $\theta$ and density function $[\Gamma(k)\theta^k]^{-1}t^{k-1}e^{-t/\theta}$.
These submodels illustrate that individuals with lower $R_i$ values experience faster recovery, while those with higher $R_i$ values die sooner.
For each subject $i$, we define the dropout time as $T_i = \min(T_i^r, T_i^d)$.
Dropout occurred if $T_i \le t_7$. 
In this scenario, the last observed measurement time is $D_i = \lceil T_i \rceil - 1 $, where $\lceil t \rceil$ denotes the ceiling function of $t$, and the type of dropout $\delta_i$ equals 1 if $T_i = T_i^r$ (recovery) and 2 otherwise (death).
If $T_i > t_7$, dropout does not occur during the observation period, in which case $D_i = t_7$ and $\delta_i = 0$. 
In the simulation, the proportion of individuals who recover is 0.2780, 0.3496, 0.3933, and 0.4390, and the proportion of individuals who die is 0.3574, 0.3171, 0.3095, and 0.3028, corresponding to $AD$ values of 0.1, 0.11, 0.12, and 0.13.

\subsubsection{Estimands} 
Consistent with the previous simulation, we aim to estimate the average distance ($AD$), a function of $c_0$, $c_1$, $d_0$ and $d_1$.

\subsubsection{Methods} 
This simulation study compares two distinct joint models with a reference model. 
All three models use the proposed folded normal mixed-effects regression for the observed magnitude outcomes.
Model I, the reference, does not model the dropout processes. 
Model II and Model III employ linear and categorical time functions, defined in equations (\ref{f8}) and (\ref{f7}) respectively, within their dropout submodels.
For each $AD$ scenario, 1000 simulation runs are conducted. All three models undergo a 2000-sample burn-in period to ensure Markov chain convergence before running four chains, each with 2000 samples.

\subsubsection{Performance Measures} 
We estimate the posterior distribution of $AD$ for all three models. 
Model performance is assessed using the average (posterior) mean, bias, relative bias, median, standard deviation, mean squared error, $2.5\%$ and $97.5\%$ quantiles, and the coverage probability of the $95\%$ credible interval. 
We also compute the empirical standard error to evaluate the estimation of the standard deviation.

\subsubsection{Results} 
Table \ref{tab:Tab7} presents the simulation results. 
For each model, the average posterior mean closely aligns with the average posterior median, indicating a non-skewed distribution for AD. 
For all three models, the empirical standard errors are somewhat smaller than the average standard deviations, indicating that these models tend to overestimate their own uncertainty.
Table \ref{tab:Tab7} also shows a negative average bias across all three models, suggesting missing data causes underestimation of the average distance ($AD$). 
However, Model I exhibits significantly larger absolute bias and absolute relative bias than Models II and III, which have comparable bias performance.
Although Model I, being the simplest, displays the smallest average standard deviation and the narrowest $95\%$ credible interval, its substantially larger bias causes its coverage probability to deviate considerably from the nominal level. 
In contrast, Models II and III achieve coverage probability within $3\%$ of the nominal level.
Finally, the average mean squared error suggests Models II and III perform similarly and both significantly outperform the reference Model I. 
Overall, our simulation demonstrates that joint models can mitigate bias from informative dropout. Furthermore, Model II is preferred as it offers similar bias reduction with a simpler model structure than Model III.

\begin{table}[ht]
	\renewcommand*{\arraystretch}{2}
	\caption{Performance comparison of joint models with linear (II) and categorical (III) time functions against the reference model (I) across four true average distance (TAD) scenarios.}
	\centering
	\setlength{\tabcolsep}{4pt} % Reduces space between columns
	\begin{tabular}{cccccccccccc}
		\hline 
		Model & TAD & Bias & Rel. Bias & Mean & Med. & S.D. & S.E. & $2.5\%$ & $97.5\%$ & C.P. & MSE \\ 
		\hline 
		\multirow{4}{1em}{I} & 0.10 & -0.01617 & -0.1617 & 0.0838 & 0.0836 & 0.0133 & 0.0127 & 0.0584 & 0.111 & 0.776 & 0.000601 \\
        & 0.11 & -0.01692 & -0.1538 & 0.0931 & 0.0929 & 0.0131 & 0.0127 & 0.0679 & 0.119 & 0.742 & 0.000621 \\
        & 0.12 & -0.01719 & -0.1433 & 0.1028 & 0.1026 & 0.0131 & 0.0127 & 0.0778 & 0.129 & 0.743 & 0.000629 \\
        & 0.13 & -0.01632 & -0.1255 & 0.1137 & 0.1134 & 0.0132 & 0.0130 & 0.0884 & 0.140 & 0.774 & 0.000613 \\
		\hline  
		\multirow{4}{1em}{II} & 0.10 & -0.00541 & -0.0541 & 0.0946 & 0.0942 & 0.0149 & 0.0138 & 0.0663 & 0.125 & 0.951 & 0.000444 \\
        & 0.11 & -0.00637 & -0.0579 & 0.1036 & 0.1032 & 0.0147 & 0.0138 & 0.0759 & 0.134 & 0.941 & 0.000447 \\
        & 0.12 & -0.00705 & -0.0588 & 0.1130 & 0.1126 & 0.0146 & 0.0138 & 0.0853 & 0.143 & 0.924 & 0.000455 \\
        & 0.13 & -0.00719 & -0.0553 & 0.1228 & 0.1224 & 0.0147 & 0.0140 & 0.0950 & 0.153 & 0.929 & 0.000466 \\
		\hline
		\multirow{4}{1em}{III} & 0.10 & -0.00530 & -0.0530 & 0.0947 & 0.0943 & 0.0149 & 0.0137 & 0.0665 & 0.125 & 0.947 & 0.000442 \\
        & 0.11 & -0.00633 & -0.0575 & 0.1037 & 0.1033 & 0.0147 & 0.0136 & 0.0759 & 0.134 & 0.943 & 0.000442 \\
        & 0.12 & -0.00710 & -0.0592 & 0.1129 & 0.1125 & 0.0146 & 0.0135 & 0.0853 & 0.143 & 0.931 & 0.000450 \\
        & 0.13 & -0.00724 & -0.0557 & 0.1228 & 0.1223 & 0.0148 & 0.0139 & 0.0949 & 0.153 & 0.932 & 0.000468 \\
		\hline          
	\end{tabular} 
	\label{tab:Tab7}
\end{table}

\section{Application}

We now apply our methods to analyze data from our motivating study, ``Evolution of Diaphragm Thickness during Mechanical Ventilation''.
This analysis includes more subjects than initially reported by \citet{goligher2015evolution} as recruitment continued beyond their initial publication. 
We excluded patients with only baseline diaphragm thickness, resulting in a dataset of 203 patients. 
Of these, 170 experienced dropout, with 36 due to deterioration leading to death and the rest due to recovery. Thickness measurements were taken at baseline, then daily from day one to day nine.

Our analysis investigates the potential association between sex (as the exposure variable) and the absolute change in diaphragm thickness. 
We classified individuals into two exposure groups based on sex: 75 females and 128 males.
Table \ref{tab:Tab1} shows the number of dropouts over time for each group, categorized by dropout type.
In this table, columns labeled $D_i$, for $1 \le i \le 8$, represent the number of dropouts occurring between day $i$ and day $i+1$, while the column labeled $D_9$ indicates the number of subjects who completed the observation period without dropout.
Notably, dropouts due to death are significantly fewer than those attributed to recovery across most time periods.
Overall, $59.19\%$ of records exhibited diaphragm thickness values smaller than baseline, with females at $64.89\%$ and males at $55.78\%$. 
The average absolute change in thickness was 0.132 for the entire population, or 0.125 for females and 0.136 for males.

\begin{table}[ht]
	\renewcommand*{\arraystretch}{2}
	\caption{Dropout distribution by sex and type over time.}
	\centering
	\begin{tabular}{ccccccccccc}
		\hline 
		Group & Type & $D_1$ & $D_2$ & $D_3$ & $D_4$ & $D_5$ & $D_6$ & $D_7$ & $D_8$ & $D_9$ \\ 
		\hline 
		\multirow{2}{3em}{female} & recovery & 12 & 8 & 9 & 1 & 3 & 5 & 5 & 5 & \multirow{2}{1em}{12}\\
         & death & 0 & 0 & 3 & 0 & 1 & 4 & 5 & 2 & \\
		\hline 
		\multirow{2}{3em}{male} & recovery & 24 & 9 & 9 & 11 & 3 & 9 & 9 & 12 & \multirow{2}{1em}{21}\\
         & death & 3 & 3 & 1 & 2 & 4 & 1 & 1 & 6 & \\
		\hline 
        
	\end{tabular} 
	\label{tab:Tab1}
\end{table}

We used four different models to fit the data: 
Model A, a standard normal linear mixed-effects model; 
Model B, a folded normal mixed-effects model without accounting for missing data due to dropouts; 
Model C, a joint model that simultaneously models the outcome and dropout processes, employing linear time functions (\ref{f8}) in its dropout submodel; 
and Model D, a joint model with categorical time functions (\ref{f7}), which was simplified to reduce computational complexity by combining adjacent measurement times into single indicator variables, thereby reducing the number of parameters by 16.

We estimated the exposure effect, the average distance ($AD$) between the regression lines of the two groups over time, for each model. 
The analysis used the Markov Chain Monte Carlo (MCMC) method, implemented with JAGS 4.3.0 in the R 4.0.2 environment. For all four models, we used a burn-in of 4,000 samples, subsequently running four chains with 4,000 samples per chain. 
We monitored the trace of each chain to ensure Markov chain convergence.

The main results are summarized in Table \ref{tab:Tab2}. 
Models C and D (joint models) show larger $AD$ values (0.00997 and 0.01169, respectively) compared to Model B (0.00840), which does not account for dropout. 
The folded normal models (B, C, D) have somewhat larger $AD$ values than Model A (linear model), which has a posterior mean of 0.00795. 
Model A also demonstrates a slightly larger standard deviation compared to the folded normal models. 
Finally, the $95\%$ credible intervals for all four models cross 0, indicating little evidence that sex modifies the change in diaphragm thickness.
In this case, since the effect of sex on the magnitude of the change in diaphragm thickness is insignificant (the regression lines for females and males are almost identical), we expect little bias in the estimation of $AD$, regardless of the model used. 
Consequently, this scenario with minimal inherent bias doesn't prominently showcase the bias reduction advantages of the folded normal and joint models.
However, our proposed methods are still preferred as they are more likely to detect exposure effect when it is significant.
%Comparing Model B, C, and D indicates that incorporating the dropout process into the modeling results in different estimates of average distance, with the joint models showing larger effects for sex, especially for the model with flexible temporal functions (Model D).
%Furthermore, the comparison between folded normal models and the linear model (Model A) confirms the idea that bias might emerge when not assuming a folded normal distribution for magnitude data.

\begin{table}[ht]
	\renewcommand*{\arraystretch}{2}
	\caption{The analysis results for average distance in the four models.}
	\centering
	\begin{tabular}{cccccc}
		\hline 
		Model & Mean & Median & S.D. & $2.5\%$ & $97.5\%$  \\ 
		\hline
  	    A & 0.00771 & 0.00785 & 0.0129 & -0.0183 & 0.0327 \\
		B & 0.00840 & 0.00819 & 0.0120 & -0.0151 & 0.0323 \\
		C & 0.00997 & 0.01003 & 0.0124 & -0.0140 & 0.0340 \\
		D & 0.01169 & 0.01174 & 0.0124 & -0.0133 & 0.0357 \\
        \hline
	\end{tabular}
	\label{tab:Tab2}
\end{table}

\section{Discussion}
Magnitude data are encountered across various fields, making the development of effective statistical models crucial. 
Motivated by a clinical study on the magnitude of change in diaphragm thickness over time, we developed Bayesian methods to model repeated measures of folded normal magnitude outcomes.

Our methods extend previous work by incorporating random effects into a folded normal regression model to account for subject-level heterogeneity. 
We addressed the key issue of non-identifiability by formulating suitable constraints for these random effects. Furthermore, we presented a joint modeling approach that addresses missing data arising from informative dropout, a common challenge in repeated measures studies. 
Our joint models utilize a discrete-time competing risk model to analyze dropout processes, offering a flexible framework for diverse dropout patterns.
Simulations confirmed the validity of our methods under scenarios that closely resemble the conditions of the motivating study, showing that the proposed folded normal mixed-effects model is essential for accurate magnitude data analysis, and our joint models effectively eliminate bias from informative dropout.

A current limitation of our approach is the inclusion of additional covariates, as the current models only accommodate treatment/exposure and measurement time. 
While sufficient for randomized clinical trials (where baseline characteristics are balanced), adjusted analyses are often preferred in observational studies. 
Future work will therefore focus on incorporating additional covariates, though this will increase the complexity of addressing non-identifiability, particularly for continuous covariates. 
Extending models for discrete covariates by assuming distinct regression lines is feasible but practical only for a limited number of categories. 
Careful consideration of these challenges is required.

Another important avenue for future research involves extending our methods to non-linear regression models. Currently, our approach assumes a linear relationship between outcome and time. 
While this sufficed for our motivating study, more complex non-linear relationships may be necessary in other contexts. 
However, applying non-linear regression (e.g., polynomial or spline) to magnitude outcomes with random effects would be challenging, as it would necessitate sophisticated constraints or techniques to ensure identifiability.

In conclusion, while different studies present varying modeling challenges for folded normal regression, the concepts underlying our methodology, motivated by a specific study, are broadly applicable. 
We have provided guidelines for Bayesian mixed-effects modeling of magnitude data and demonstrated that the proposed methods perform well in terms of estimation accuracy. 
Overall, we have effectively addressed the challenges of analyzing magnitude data with informative dropout, particularly those posed by our motivating study.

\bibliographystyle{plainnat}
\bibliography{references}  %%% Uncomment this line and comment out the ``thebibliography'' section below to use the external .bib file (using bibtex) .





% \end{thebibliography}

\end{document}